          [2001/04/25 1.1 (PWD)]
\documentclass[a4paper,twocolumn]{esapub} % European paper
\usepackage{natbib}
\usepackage{amsmath}
\usepackage{epsfig}
\usepackage{graphicx}
\usepackage{rotate}
\title{Coordinated INTEGRAL and optical observations of SS433}
\author[(1)]{A. M. Cherepashchuk}
\author[(2)]{R. A. Sunyaev}
\author[(3)]{S. N. Fabrika}
\author[(2)]{S. V. Molkov}
\author[(3)]{E. A. Barsukova}
\author[(1)]{E. A. Antokhina}
\author[(1)]{T. R. Irsmambetova}
\author[(1)]{I. E. Panchenko} 
\author[(1)]{K. A. Postnov}
\author[(1)]{E. V. Seifina}
\author[(1)]{N. I. Shakura}
\author[(1)]{A. N. Timokhin}
\author[(4)]{I. F. Bikmaev}
\author[(4)]{N. A. Sakhibullin}
\author[(5)]{Yu. N. Gnedin}
\author[(5)]{A. A. Arkharov}
\author[(6)]{V. M. Larionov}

\affil[(1)]{\it Sternberg Astronomical Institute, 119992 Moscow, Russia,
cher@sai.msu.ru}
\affil[(2)]{\it Space Research Institute, Moscow, Russia,
molkov@hea.iki.rssi.ru}
\affil[(3)]{\it Special Astrophysical Observatory, Nizhny Arkhyz,
Karachaevo-Cherkesiya, Russia, fabrika@sao.ru}
\affil[(4)]{\it Kazan State University, Kazan, Tatarstan, Russia,
Ilfan.Bikmaev@ksu.ru}
\affil[(5)]{\it Pulkovo Observatory, Sanct-Petersburg, Russia, gnedin@gao.spb.ru}
\affil[(6)] {\it Astronomical Institute of
Sanct-Petersburg State University, Russia, vlar@nm.ru}

\begin{document}
\maketitle


\begin{abstract}
Results of simultaneous {\it INTEGRAL} and optical observations 
of galactic microquasar SS433 in May 2003 are
presented. The analysis of the X-ray and optical eclipse duration
and hard X-ray spectra obtained by {\it INTEGRAL} together with
optical spectroscopy obtained on the 6-m telescope allows us to construct a 
model of SS433 as a massive X-ray binary.
X-ray eclipse in hard X-rays has a depth of $\sim 80\%$ and
extended wings. The optical spectroscopy allows us to identify
the optical companion as a A5-A7 supergiant and to measure its radial
velocity semi-amplitude $K_v=132$ km/s. A strong heating effect in the 
optical star atmosphere is discovered spectroscopically. 
The observed broadband  X-ray spectrum 2-100 keV can be described by
emission from optically thin thermal plasma with $kT\sim 15-20 keV$    
\end{abstract}

\section{Introduction}

SS433 is a supercritically accreting microquasar with precessing 
accretion disk and collimated mildly relativistic ($v\approx 0.26 c$ ) jets
(see e.g. Cherepashchuk et al. 1996 and references therein). 
Since the discovery in 1978, this unique massive X-ray 
binary system has been investigated in optical, radio and X-rays 
by many authors (e.g.
% Clark and Murdin
%1978,  Margon et al. 1980, Mammano et al. 1980, Milgrom 1979, 
%Fabian and Rees 1979, Crampton et al. 1980, van den Heuvel et al. 1980, 
Cherepashchuk 1981,  
%Crampton and Hutchings 1981, 
Margon 1984, 
%Cherepashchuk 1988, Cladyshev et al. 1987, Kemp et al. 1987, Stewart et al. 1987,
%Vermeulen et al. 1993, 
Kawai et al. 1989, Kotani et al. 1996, 
%Kotani 1998, 
Goranskii et al. 1998a,b, 
%Panferov and Fabrika 1997, 
Marshall et al 2002, Fabrika 2004). 

SS433 is highly variable system and shows different types of
periodicities:

1. Precessional variability ($P_{prec}=162^d.5$) which is observed by 
periodical Doppler-shifts of H, He I, Fe XXV and other X-ray emission lines
in optical and X-ray spectra and is clearly visible 
in optical and X-ray light curves. 

2. Orbital periodicity ($P_{orb}=13^d.082$). The shape of the orbital
optical
light curve strongly changes with precession phase (Goranskii et al. 1998
a,b, 
Cherepashchuk and Yarikov 1991). The value of the orbital period remains
stable over more than 30 years, which can be considered as an argument
against the common envelope model for SS433. 

3. Nutation periodicity ($P_{nut}=6^d.2877$) which is observed as periodic 
deviations from purely precessional Doppler motion of emission lines and 
can be recovered from photometric data (Goranskii et al. 1988   
a,b, 
Cherepashchuk and Yarikov 1991). The period and phase of the nutational
variability remain stable over at least 16 years (around 950 nutation periods). 
Nutation radial velocity variations are delayed from the nutation photometric 
variability  by 0.6 days, which corresponds to the travel time from the
accretion disk center to the formation region of optical moving emission
lines downstream the relativistic jets located at $l\simeq 10^{14}-10^{15}$ cm away
from the disk center.

The AO-1 {\it INTEGRAL} observations of SS433 discovered hard (up to 100 keV)
X-ray spectrum in this supercritically accreting microquasar (Cherepashchuk
et al. 2003), suggesting the presence of an extended hot (with temperature
up to $10^8$ K) region at the central parts of the accretion disk. These new
data made it possible to compare the eclipse characteristics of SS433 at
different energies: soft X-rays (2-10 keV, the {\it ASCA}
data), soft and medium X-ray (1-35 keV, the {\it Ginga} data),
hard X-ray (20-70 keV, the {\it INTEGRAL} data), and optical. Such a comparison
allows us to study the innermost structure of the supercritical accretion
disk and to constrain the basic parameters of the binary system. This is
especially important in view of recent controversial results on 
the optical spectroscopy of
SS433 (Gies et al. 2002, Charles et al. 2004).
 
\section{Observational campaign}

\begin{figure}
\centering
\epsfig{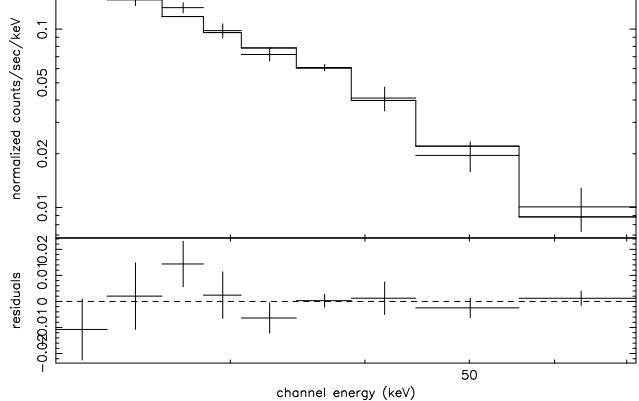}
\caption{IBIS/ISGRI spectrum of SS433 collected over {\it INTEGRAL}
orbits 67-69 in May 2003. The best fit (solid line) is for exponential
cut-off with $kT=14\pm 2$ keV (reduced $\chi^2=0.32$ for 7 dof, 
$CL\approx 95\%$).
\label{fig:IBIS_sp}}
\end{figure}

\begin{figure}
\centering
\epsfig{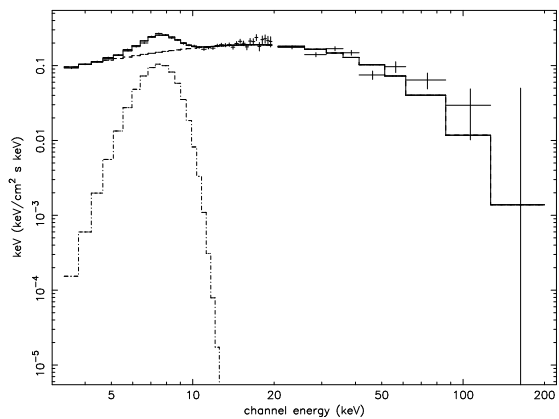}
\caption{RXTE PCA and IBIS/ISGRI spectrum of SS433
obtained during simultaneous {\it RXTE/INTEGRAL} observations
of SS433 in March 2004 (JD 2453076-2453078). 
The best fit (solid line) is for emission from optically 
thin plasma with $kT=19.5\pm 1 keV$.  
%(reduced $\chi^2=0.63$ for 6 dof).
\label{fig:RXTEIBIS_sp}}
\end{figure}

\begin{figure}
\centering
\epsfig{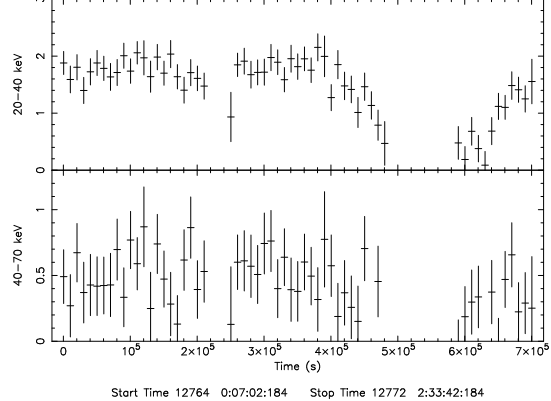}
\caption{IBIS/ISGRI 20-40 keV (upper panel) and 40-70 keV
(bottom panel) count rates of SS433 (without background subtraction).
The egress part of the X-ray eclipse was kindly provided by 
Diana Hannikainen.  
\label{fig:IBIS_lc}}
\end{figure}

The coordinated multiwavelength observational campaign of SS433 was 
organized during the AO-1 {\it INTEGRAL} observations of the SS433 field in 
March-June 2003 with the participation of the following research  
teams.

1. Special Astrophysical Observatory of Russian Academy of Science
(SAO RAS) 
(S. Fabrika, E. Barsukova). Provided high signal-to-noise optical 
spectroscopy of SS433 at the 6-m telescope. 

2. Crimean Station of Sternberg Astronomical Institute (V. Lyuty, 
T. Irsmambetova). Made BVR-photometry at the 0.6-m telescope. 

3. Kazan State University. Performed optical photometry
at the Russian-Turkish  
1.5-m telescope RTT-150 at the TUBITAK National Observatory, Turkey 
(I. Bikmaev, N. Sakhibullin).

4. Infrared K-photometry was performed at 1.1-m AZT-24 
telescope of Pulkovo
Observatory in Italy, Observatore di Campo Imperatore (Yu. Gnedin,
A. Arkharov, V. Larionov). 

5. Radio monitoring at cm wavelength has been performed at the 
RATAN-600 radio telescope of SAO RAS, indicating
the source to be at its non-flaring state 
(see Trushkin 2003 for more detail). 

The results of a quick-look analysis of the {\it INTEGRAL} observations of SS433
were published earlier (Cherepashchuk et al. 2003). Here we present the
results of more 
detailed study of the {\it INTEGRAL} data together with the simultaneous optical
spectroscopic and photometric observations of SS433. More complete 
studies of all data on SS433 obtained during this campaign 
will be reported elsewhere.

\begin{figure}
\centering   
%\vspace{4cm}
%\includegraphics[width=0.8\linewidth,rotate]{orbits67_68_69.ps}
\epsfig{file=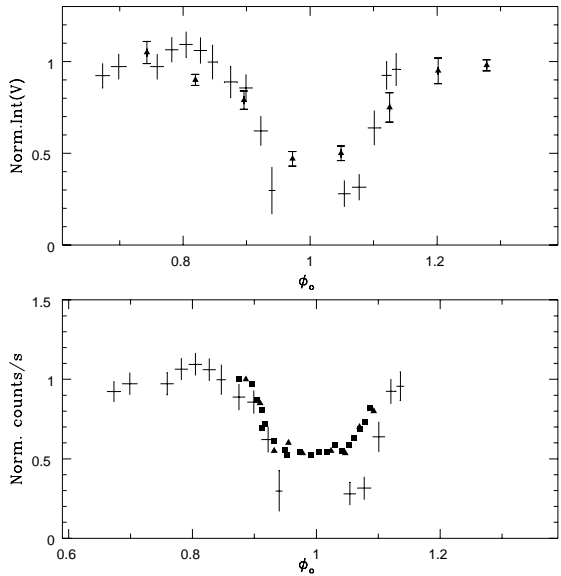,width=\linewidth}
\caption{Upper panel: Averaged over 20 ks 
IBIS/ISGRI 25-50 keV 
eclipse light curve superimposed on the mean V-light curve of SS433
(filled triangles) 
simultaneously obtained at the SAI Crimean Laboratory 0.6-m telescope, 
normalized to the maximum flux. Bottom panel: 
The same IBIS/ISGRI light curve and normalized to the maximum 
flux {\it Ginga} 5-28 keV light curve (filled squares; from Kawai et al.
1989) 
and {\it ASCA} 6-10 KeV light curve (filled triangles; from Kotani et al. 
1996).   
\label{fig:INTandV}}
\end{figure}

\begin{figure}
\centering   
\epsfig{file=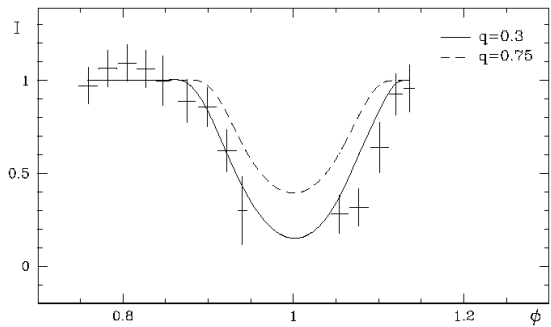,width=\linewidth}
\caption{Best fit of the IBIS/ISGRI X-ray eclipse light curve 
by the binary system model with the 
mass ratio $q=m_x/m_v=0.3$ (the solid curve)
and q=0.75 (the dashed curve). Hard X-ray emission is modeled 
by a semi-ellipsoid with axes $a_j=0.3a$, $b_j=0.75a$ posed at the 
center of a thick disk with radius $r_d=0.47a$ (for the case $q=0.75$) 
and $a_j=0.3a$, $b_j=0.75a$, $r_d=0.37a$ (for $q=0.3$).
Here $a\approx 85 R_\odot$ (for $q=0.3$) is the binary semi-major axis.  
\label{fig:q_model}}
\end{figure}

\section{IBIS/ISGRI light curves and spectra}

The {\it INTEGRAL} IBIS/ISGRI data were analyzed using  
publically available ISDC software (OSA-3 version). 
The main results are as follows.

3.1. The IBIS/ISGRI spectrum of SS433 (25-100 keV) obtained in May 2003 
can be equally well fitted both by a
power-law $\sim E^{-\alpha}$ with photon index $\alpha\approx 2.7\pm 0.13$
or optically thin thermal plasma emission with $kT\sim 12-16$ keV 
(Fig. \ref{fig:IBIS_sp}). The source was not significantly 
detected by the Jem-X telescope during these observations.
However, we made use of TOO {\it RXTE} observations of 
SS433 performed simultaneously with {\it INTEGRAL} 
in the end of March 2004 (PI: M.Revnivtsev) to 
obtain the 2-100 keV spectrum of SS433. The source was observed
about one binary orbit ($\sim 14$ days)  
later than the disk maximum opening phase. The resulting
spectrum is shown in Fig. \ref{fig:RXTEIBIS_sp} and is fitted by 
emission from optically thin thermal plasma with $kT\sim 20$ keV.
 
%Spectral fitting obtained from the analysis of IBIS/ISGRI images
%in different energy channels yields the similar value $\alpha\sim 2.9$.
%Analysis of public RXTE HEXTE data confirmed the
%absence of the high-energy cut-off in the S433 spectrum up to 60 keV.
%The obtained spectrum is significantly steeper
%than the hard X-ray power-law tails observed in subcritically accreting
%black hole transients. Applying the standard Comptonization theory
%for such a power-law spectrum and assuming the cut-off energy above 60
%keV (i.e. $kT\sim 40-60$ keV for scattering medium), 
%the optical depth of the scattering medium is $\tau \sim 0.3????$.
%This corresponds to the Comptonization parameter $y\sim 0.003???$. 
%Such a medium
%can represent a non-thermal rarefied wind around thermal X-ray 
%jets in the innermost region of the supercritical accretion disk.

The integrated hard X-ray luminosity is 
$L_x(18-60 \hbox{keV})\sim 4\times 10^{35}$
erg/s, $L_x(60-120 \hbox{keV})\sim 2\times 10^{35}$ erg/s 
(assuming the 5 kpc distance to SS433), which is about 10\% of the soft
X-ray jet luminosity.

3.2. No rapid variability was found in the hard
X-ray band on characteristic timescales $\le 1$ hour. 

3.3. 
The IBIS/ISGRI count rates of SS433 are presented in Fig. \ref{fig:IBIS_lc}.
The X-ray eclipse at hard energies is observed to 
be a little narrower than the optical one, slightly broader than in the 1-35 keV
energy range and display extended wings 
(Fig. \ref{fig:INTandV}). This is opposite to what is found in 
ordinary eclipsing X-ray binaries (like Cen X-3, Vela X-1 etc.), in which
the X-ray eclipse duration decreases with energy. This new fact may
reflect a complicated structure of the innermost supecritical accretion
disk in SS433. 

3.4. The eclipse depth is about 80\% in hard X-rays compared to 
$\sim 50\%$ in 6-10 and 5-28 keV band (Fig. \ref{fig:INTandV}).  

3.5. The 25-50 keV X-ray flux increases from $\sim 5$ to $\sim 17$ mCrab 
during precession period in Marh-May 2003. This modulation is 1.5-2 times
larger than observed in 2-10 keV egergy band. Thus, both precessional and 
eclipsing hard X-ray variabilities in SS433 exceed by 1.5-2 times those  
in the standard X-ray band. This suggests a more compact hard X-ray emitting
region in the central parts of the accretion disk. 

The observed X-ray eclipse was interpreted by the model of a close binary
with thick precessing accretion disk (Antokhina et al. 1992). The model 
includes the following basic elements: 
i) Dark optical star filling its Roche lobe. ii) Dark thick precessing
accretion disk with a central cone-like region (the cone angle $\sim 60^o$). 
iii) Bright extended bulge located at the central cone of the accretion disk. 
The observed X-ray eclipse in the 25-50 keV band is best fitted by this model
for the mass ratio $q=m_x/m_v\simeq 0.2$ and the bright bulge size 
$\sim 0.3 a$, where $a\simeq 85 R_\odot$ (for $q=0.3$) 
is the binary semi-major axis (Fig. \ref{fig:q_model}).
However, for reasons discussed below we shall assume $q=0.3$ as a 
fiducial value.

\section{Optical observations}

\begin{figure}
\centering
\epsfig{file=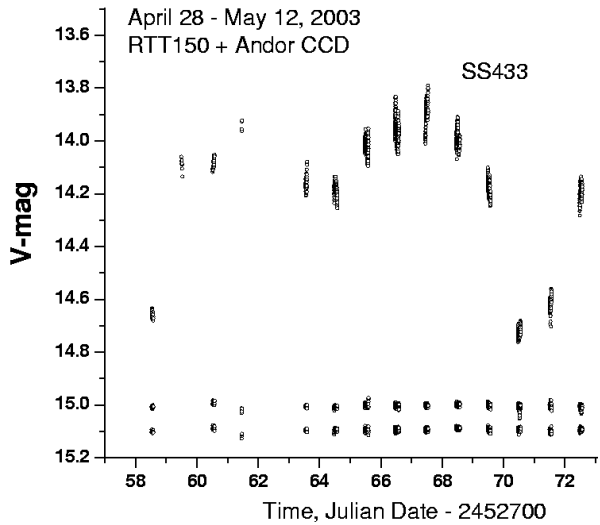,width=\linewidth,angle=180}
\caption{V-light curve of SS 433 
obtained at the RTT150 telescope (TUBITAK National Observatory, Turkey)
simultaneously with {\it INTEGRAL} observations. In the bottom: photometry
of control stars
($V_{N3}=12.975\pm 0.006, V_{N4}=12.746\pm 0.007$).
\label{fig:Kazan}}
\end{figure}

\subsection{Optical photometry}

Photometry of SS433 was performed simultaneously with {\it INTEGRAL}
observations by the Russian-Turkish RTT-150 telescope of Kazan University at
the TUBITAK National Observatory, Turkey. 
Observations were made by using commercial,
Thermoelectrically cooled to -60 C  ANDOR firm CCD
(model DW436, www.andor-tech.com) provided for
RTT150 by MPA. Low noise
back-illuminated EEV chip has 2048 x 2048 pixels
of 13.5 mkm size each.
Full field of view is 8 x 8 arcmin with frame reading
time of 40 sec at 2 x 2 binning. To increase the time
resolution  only parts of the field of view with reference
stars N 1,2,3,4,5 ( Leibowits and Mendelson, 1982) around
SS433 were stored to PC.
The obtained V-light curve is presented in Fig.
\ref{fig:Kazan}. Strong ($\sim 0.15$ mag) intranight variability of the
source on timescales $\sim 100$ s - 100 min is clearly detected. The mean V light
curve during the eclipse obtained at the Crimean Laboratory of SAI 
simultaneously with the {\it INTEGRAL} observations is also
shown in Fig. \ref{fig:INTandV}. The optical eclipse minimum 
is observed at $JD=2452770.863$, as predicted by the orbital ephemeris
given by Goranskii et al. 1998a.

\subsection{IR-photometry}

\begin{figure}
\centering
\epsfig{file=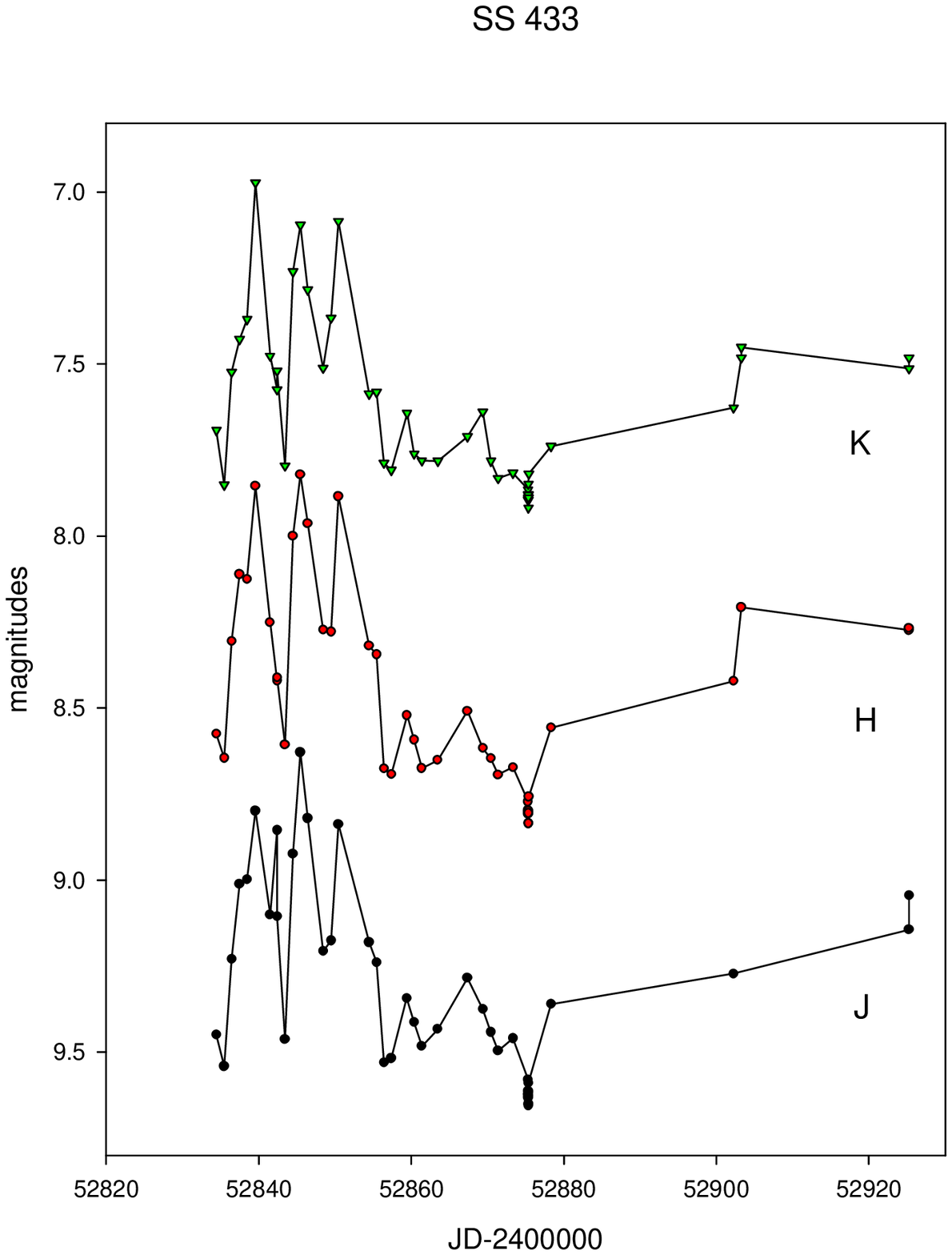,width=\linewidth}
\caption{JHK-photometrical light curve of SS433 
obtained by AZT-24 1.1-m 
IR telescope in July-August 2003 (Campo Imperatore, Italy).
\label{fig:gnedin}}
\end{figure}

Near IR observations of SS433 were obtained at the AZT-24 1.1m telescope in
Campo Imperatore (Italy) with SWIRCAM during a period spanning July-August
2003. SWIRCAM is the infrared camera that incorporates a 256x256 HgCdTe
NIGMOS 3-class (PICNIC) detector at the focus of AZT-24. It yields a scale
of 1".04/pixel resulting in a field of view of ~ 4x4 sq. arcmin. The
observations were performed through standard JHK broadband filters.
The NIR monitoring of SS433 started several orbits after 
the {\it INTEGRAL} observations. 
The JHK light curves obtained in July-August 2003
are presented in Fig. \ref{fig:gnedin}.
The observations were done close to the cross-over precessional phase, 
where the orbital modulation appears to be significantly reduced. 
%The primary IR minima appear to be slightly off-phase with respect to 
%the expected moments of the optical minima, 
%which may suggest an asymmetric shape of the wind from 
%the accretion disk and the optical star. 

\subsection{Optical spectroscopy}

\begin{figure*}
\centering
\epsfig{file=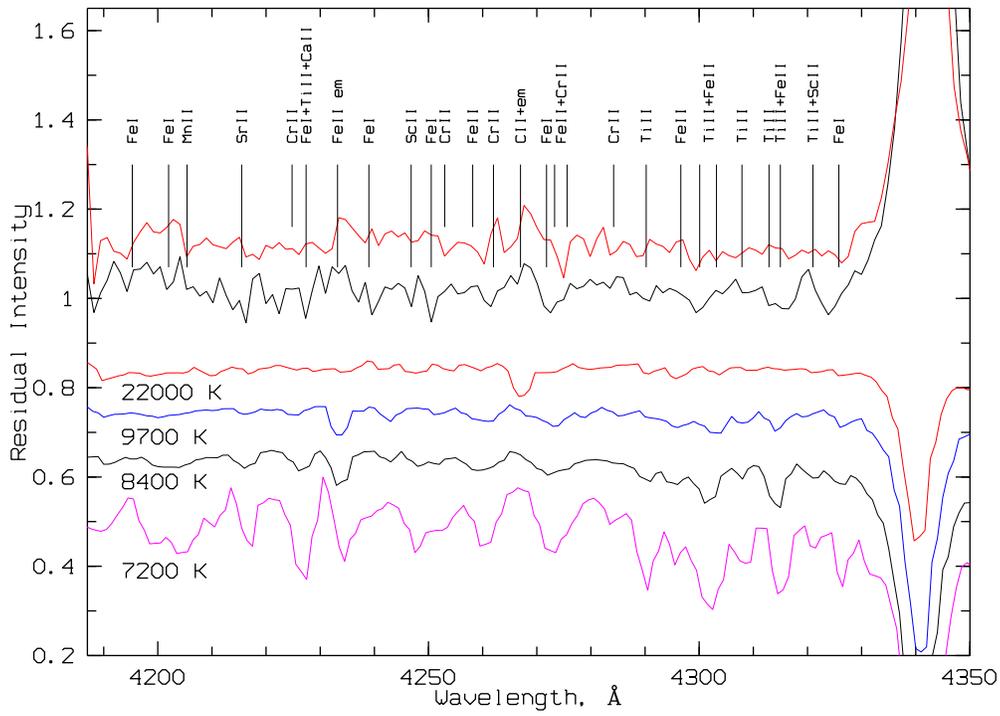,width=0.6\textwidth,angle=-90}
\caption{Optical spectra of SS433 taken on the 6-m telescope
and spectra of supergiants of known temperatures (Le Borgne et al. 2003)
for a comparison.
Two upper spectra are
averaged over nights 28/04/03 and 11/05/03
(the disk eclipse phase) (second from the top
spectrum in the figure) and over 09/05/03 and 12/05/03 (corresponding to
evenly spaced from the eclipse center orbital phases 0.88 and 0.12) (the top
spectrum).
\label{fig:SPklass}}
\end{figure*}

\begin{figure*}
\centering   
\epsfig{file=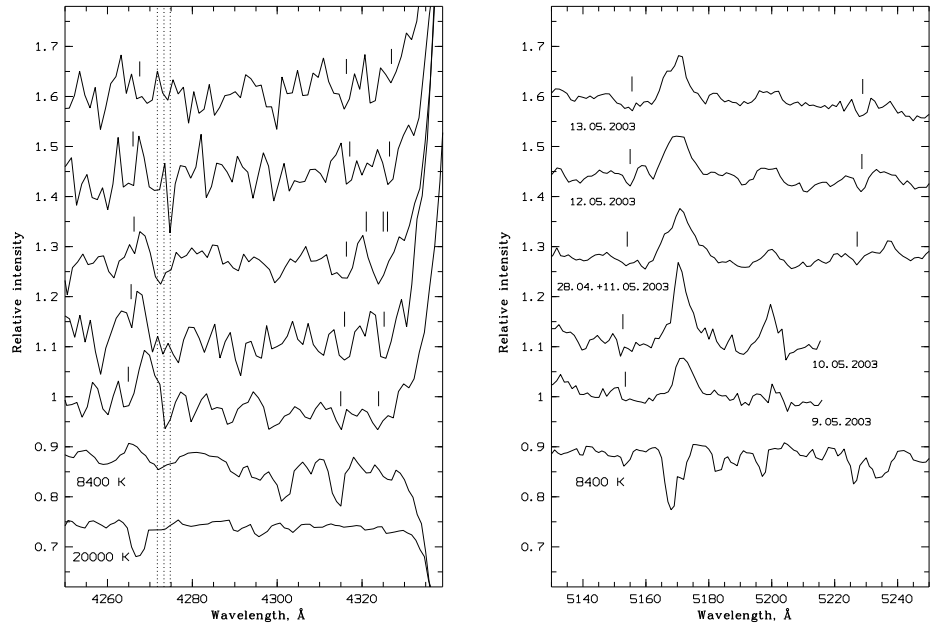,width=0.6\textwidth,angle=-90}
\caption{Evolution of absorption lines in the 
optical spectra of SS433 taken at different orbital phases.
For comparison, spectra of 8400 K and 20000 K supergiants are shown
in the bottom. Orbital evolution of blends and radial
velocities are traced by different absorption lines (dashes).
See the text for more detail. 
\label{fig:demo21}}
\end{figure*}

The optical spectroscopy of SS433 was carried out 
at the 6-m telescope of SAO RAS
during 6 nights in April 28 and May 9-13 2003, simultaneously with the
{\it INTEGRAL} observations, at the precessional phase corresponding to the
maximum disk opening angle. The long-slit spectrograph and PM-CCD 1000x1000
detector were used to obtain 4100-5300 A spectra with a resolution of 3~A
(1.2 A/pix). The blue part of the optical spectrum including He II 4686
emission line was chosen for the analysis. The standard technique was used
for spectral reduction and calibration. Fig. \ref{fig:SPklass} shows four
spectra of supergiants with known temperatures (Le Borgne et al. 2003) and
SS433 spectra averaged over nights 28/04/03 and 11/05/03 (the disk eclipse
phase) (second from the top spectrum in the figure) and over 09/05/03 and
12/05/03 (corresponding to evenly spaced from the eclipse center orbital
phases 0.88 and 0.12) (the top spectrum). Spectral resolution in SS433
spectra and in those of standard supergiants is the same. The
signal-to-noise ratio in our spectra is $>60$ at $\lambda 4250$ A 
per resolution element. Fe~II 4233 emission is seen, emission lines in SS433 spectrum are broad
($FWHM\sim 10 A$). CII 4267 and NIII 4196,4200 + HeII 4200 emissions are
also marginally present. In the Figure, the absorption lines seen out of the
eclipse are marked with short vertical bars; those which are seen in the
eclipse or present in both spectra are marked with long bars.

The intensity of absorption lines during the eclipse allows us to 
estimate the effective temperature of the optical star to be $T<9000$ K.
The relative intensities of the strongest absorption lines indicate
$T=8000\pm 500$ K, implying the optical spectral class of the companion 
A5-A7I. 

The heating effect of the companion atmosphere is discovered by
absorption lines. During the disk eclipse egress, low-excitation 
absorption lines strongly weaken. The stellar hemisphere illuminated by
the bright accretion disk probably has a temperature of $\sim 20,000$ K, 
as the presence of CII 4267 absorption suggests. 
%%%%%% 
This is the only strong line in this region in hot stars. The CII 4267
absorption is deep out of the eclipse (Fig. \ref{fig:SPklass}) and the line
is only marginally detected inside the eclipse phases.    
Evolution of
the blend 4273 (FeI 4271.8 + FeII 4273.3 + CrII 4275.6) is seen: in the
eclipse, low-excitation line FeI is strongest, while out of eclipse
FeII+CrII lines enhance. Also, other FeI lines appear only in the
eclipse. These effects are illustrated 
%%%%%%
better by Fig. \ref{fig:demo21}, in which 
spectra of standard stars with different effective temperatures are shown
together with spectra of SS433 obtained in 9/05, 10/05, 28/04+11/05, 
12/05 and 13/05 (orbital eclipses fell in nights 10/05 and 28/04+11/05).

\begin{figure}
\centering   
\epsfig{file=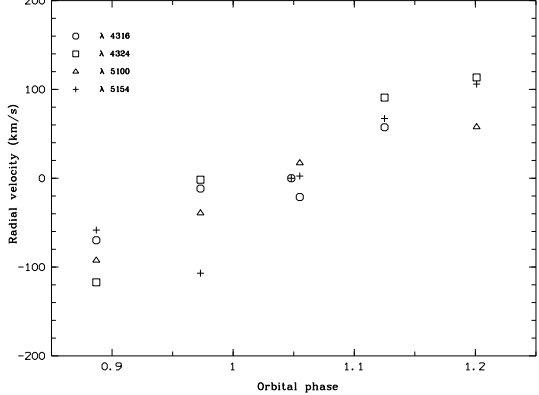,width=0.7\linewidth,angle=-90}
\caption{Radial velocity curve of the optical companion of SS433
obtained from 4 individual absorption lines over 6 nights.
\label{fig:rvindid}}
\end{figure}

\begin{figure}
\centering    
%\vspace{4cm} 
%\includegraphics[width=0.8\linewidth,rotate]{orbits67_68_69.ps}
\epsfig{file=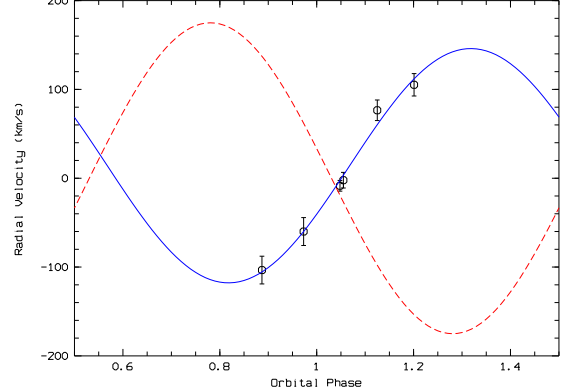,width=0.7\linewidth,angle=-90}
\caption{Mean radial velocity curve of the optical companion of SS433
measured from 22 individual absorption lines. The accretion disk 
radial velocities as measured from by HeII 4686 emission 
(Fabrika and Bychkova, 1990) is shown by the dotted line.   
\label{fig:radvel}}
\end{figure}

In Fig. \ref{fig:rvindid} we present individual 
absorption line radial velocities for four best-suited lines 
averaged over one night, 
as a function of the orbital phase for 6 nights. 
It is important to note that the strongest 
%%%%%% 
lines (FeII, HeI, hydrogen) are present in SS433 spectrum as emissions. 
Some other strong lines (among FeII and HeI lines) appear in SS433 
as P\,Cyg-like absorption components or distorted  absorptions, they   
are disfavored for the optical star radial velocity analysis, as they
most probably form in the powerful outflowing disk wind (Fabrika et al.
1997). 

The radial velocity curve measured by 
the most reliable collection of 
22 absorption lines in the spectral range 4200-5300 A
is shown in Fig. \ref{fig:radvel}. 
%%%%%%
Note that the absorption lines in SS433 spectrum are seen at our spectral
resolution as blends containing 1-3 strong lines. Relative intensities 
of the lines are probably change from date to date through the eclipse.
In this reason we measured only those lines and in only those dates 
where the lines were detected the most convincingly. The radial velocity 
(Fig. \ref{fig:radvel}) has been obtained by coadding of radial velocity
curves of individual lines and a radial velocity of all lines for 11/05/03
(the orbital phase 0.048) was adopted to be a zero. 
The derived 
radial velocity semi-amplitude of the optical star is $K_v=132\pm 9$~km/s,
the gamma-velocity of the binary system is $v_\gamma=14\pm 2$ km/s. 
The absorption line radial velocity transition through 
the $\gamma$-velocity occurs at the middle of the optical eclipse
%%%%%%
($\phi_b=0.07$), confirming that the lines actually 
belong to the optical star. 

These results confirm the earlier determination of $K_v$
by Gies et al. (2002) also made at the maximum disk opening angles
(note that 
spectroscopic observations by Charles et al. 2004 were performed at the
cross-over phase of SS433 when the accretion disk is seen edge-on; such a
phase is disfavored for the optical star 
radial velocity analysis as strong gas outflows are present 
in the disk orbital plane; selective absorption in this moving gas affects 
the true radial velocity of the optical star). 

The heating effect of the optical star also  
distorts the radial velocity semi-amplitude. The analysis 
(Wade and Horne 1988, Antokhina et al. 2004) indicates that true value of the 
semi-amplitude of the radial velocity curve as derived from these
absorption lines can be reduced to $\sim 70-80$ km/s.

\section{Basic parameters of the binary system}

Comparison of radial velocities of the accretion disk ($K_x=175$ km/s, 
Fabrika and Bychkova 1990) and optical star ($K_v=132$ km/s) yields the
mass ratio in the SS433 system $q=m_x/m_v=K_v/K_x=0.75$. This is 
an upper limit, as the true value of $K_v$ should be corrected for
the heating effect. Taken at face value, these $q, K_v$ 
lead to the optical star
mass function $f_v\approx 3.12 M_\odot$ and the binary component 
masses $m_x=18 M_\odot$, $m_v=24 M_\odot$. However, such a big mass ratio
is in a strong disagreement with the observed duration of X-ray 
eclipse, which suggests much smaller mass ratio $q\sim 0.2-0.3$. 
We stress that the binary inclination angle in SS433 ($i=78^o.8$) 
is fixed from the analysis of moving emission lines. 

There are two possibilities: (1) either the model we used to fit  X-ray 
eclipses should be modified, or (2) the value of $K_x$ and $K_v$ 
are influenced by additional physical effects. 
Though there are some reasons to modify the model (e.g., 
asymmetric shape of the X-ray eclipse, which may suggest an asymmetric wind
outflow
from the illuminated part of the optical star), we shall consider here only 
the second possibility. We repeat again that 
the actual value of $K_v$ should be 
decreased to account for the observed heating effect.  

Let us assume the mass ratio in the system to be $q=0.3$, as 
with this value we can satisfactorily describe the X-ray eclipse width  
(see Fig. \ref{fig:q_model}). Taking $K_v=132$ km/s yields
$f_v\approx 3.1 M_\odot$ and $m_x=62 M_\odot$, $m_v=206 M_\odot$, 
$K_x=440$ km/s. Clearly, this is an unacceptable model. 
Now let us decrease $K_v$ down to 85 km/s, the lower limit which follows
from more accurate treatment of the heating effect in the radial velocity
curve analysis (Wade and Horne 1988, Antokhina et al. 2004). This
would yield a better fit with $m_x=17 M_\odot$, $m_v=55 M_\odot$, and
$K_x=283$ km/s, still too high to be acceptable. Finally, let us take into
account that less than half of the stellar surface is sufficiently cool to
give the absorption lines under study (e.g. due to sideway heating from
X-ray jets and scattered UV radiation in the strong accretion disk wind).
This additionally decrease the value of the actual radial velocity
semi-amplitude. So taking, for example, $K_v=70$ km/s and $q=0.3$ yields the
optical star mass function $f_v=0.46 M_\odot$ and binary masses
$m_x=9 M_\odot$, $m_v=30 M_\odot$ and optical star radius $R_v\sim 40
R_\odot$. This radius is compatible with typical bolometric  
luminosity of a $30 M_\odot$ A5-A7 supergiant with $T_{eff}\sim 8500$~K.
In this solution, $K_x\simeq 233$ km/s, larger than the measured value 175
km/s, but the true value of $K_x$ may be affected by the strong accretion
disk wind.

\section{Conclusions}

The {\it INTEGRAL} and optical observations of SS433 provide 
support that SS433 is a massive black hole X-ray binary
system. The presence of broad wings in the 
observed hard X-ray eclipse profile evidences for an extended central
hot region in the SS433 accretion disk. This region 
is observed to be 
substantially eclipsed during precessional motion of the accretion disk.
The detected broad-band (2-100 keV) X-ray spectrum is best-fitteed by
thermal plasma with $kT\sim 15-20$~keV. The origin of such an extended
hot region in the cenral parts of supercritically accretion disk in SS433
needs more studies.     

The measured semi-amplitude of the radial velocity of the 
optical star is $K_v=132\pm 9$ km/s, which confirms the results by Gies et
al. (2002). However, the actual semi-amplitude should be decreased
down to $\sim 70$ km/s to be in agreement with the 
observed duration of X-ray and optical eclipses. This correction 
seems indeed inevitable because of strong heating effect present in the 
optical star of SS433. 

To measure more precisely the binary parameters of SS433,
further hard X-ray observations 
of SS433 are required
at different precession phases and high-resolution optical spectroscopy 
need to be carried out at the maximum disk open angle to accurately 
measure $K_v$ and $K_x$.
Analysis 
of X-ray observations at minimum X-ray flux and the cross-over 
precession phase 
will allow more detailed reconstruction of X-ray emitting zone 
in the center of the supercritically accretion disk in SS433.    

\section*{Acknowledgments}

The results presented in this paper are based on observations with 
{\it INTEGRAL}, an ESA project with instruments and science data centre funded
by ESA member states (especially the PI countries: Denmark, France, 
Germany, Italy, Switzerland, Spain), Czech Republic and Poland, and with
the participation of Russia and the USA.

The authors acknowledge E.K. Sheffer, S.A. Trushkin, 
V.M. Lyuty, V.P. Goranskii, A.A. Lutovinov
for discussions and collaboration. 
We especially thank Dr. D. Hannikainen for providing us
with data on SS433 X-ray egress observations and M.G. Revnivtsev
for processing {\it RXTE} spectra of SS433.  
IFB thanks the Turkish National Observatory staff (Prof. Zeki Aslan, Dr. 
Irek Khamitov, night assistants
Kadir Uluc and Murat Parmaksizoglu) for their support 
in the photometric SS433 observations.
We thank A.G. Pramsky and A.N. Burenkov for a help in spectral observations.
AMCh and KAP acknowledge the
financial support from ESA. 
The work of SNF 
is partially supported by the RFBR grant
04-02-16349.
The work of KAP is partially supported by the RFBR 
grant 03-02-16068. IFB and SVM are grateful to the International Space
Science Institute (Bern, Switzerland) for hospitality
and partial financial support.

\end{document}